\documentclass{article}%
\usepackage{amssymb}
\usepackage{amsmath}
\usepackage{amsfonts}
\usepackage{graphicx}%
\providecommand{\U}[1]{\protect\rule{.1in}{.1in}}
\newtheorem{theorem}{Theorem}[section]

\newtheorem{corollary}[theorem]{Corollary}

\newtheorem{proposition}[theorem]{Proposition}
\newtheorem{remark}[theorem]{Remark}

\newenvironment{proof}[1][Proof]{\noindent\textbf{#1.} }{\ \rule{0.5em}{0.5em}}
\begin{document}

\author{Vadim E. Levit\\Ariel University Center of Samaria, Ariel, Israel\\levitv@ariel.ac.il
\and Eugen Mandrescu\\Holon Institute of Technology, Holon, Israel\\eugen\_m@hit.ac.il}
\title{A characterization of K\"{o}nig--Egerv\'{a}ry graphs using a common property
of all maximum matchings}
\date{}
\maketitle

\begin{abstract}
The \textit{independence number} of a graph $G$, denoted by $\alpha(G)$, is
the cardinality of an independent set of maximum size in $G$, while $\mu(G)$
is the size of a maximum matching in $G$, i.e., its \textit{matching number}.
$G$ is a \textit{K\"{o}nig--Egerv\'{a}ry graph} if its order equals
$\alpha(G)+\mu(G)$. In this paper we give a new characterization of
K\"{o}nig--Egerv\'{a}ry graphs. We also deduce some properties of vertices
belonging to all maximum independent sets of a K\"{o}nig--Egerv\'{a}ry graph.

\textbf{Key words:} maximum independent set, maximum matching, core of a
graph, critical vertex.

\end{abstract}

\section{Introduction}

Throughout this paper $G=(V,E)$ is a simple (i.e., a finite, undirected,
loopless and without multiple edges) graph with vertex set $V=V(G)$, edge set
$E=E(G)$, and order $n(G)=|V(G)|$.

If $X\subset V$, then $G[X]$ is the subgraph of $G$ spanned by $X$. By $G-W$
we mean the subgraph $G[V-W]$, if $W\subset V(G)$. For $F\subset E(G)$, by
$G-F$ we denote the partial subgraph of $G$ obtained by deleting the edges of
$F$, and we use $G-e$, if $W$ $=\{e\}$.

If $A,B$ $\subset V$ and $A\cap B=\emptyset$, then $(A,B)$ stands for the set
\[
\{e=ab:a\in A,b\in B,e\in E\}.
\]
The neighborhood of a vertex $v\in V$ is the set
\[
N(v)=\{w:w\in V,vw\in E\},
\]
and $N(A)=\cup\{N(v):v\in A\}$, while $N[A]=A\cup N(A)$ for $A\subset V$.

By $P_{n},C_{n},K_{n}$ we mean the chordless path on $n\geq3$, the chordless
cycle on $n\geq$ $4$ vertices, and respectively the complete graph on $n\geq1$ vertices.

A set $S$ of vertices is \textit{independent} if no two vertices from $S$ are
adjacent. An independent set of maximum size will be referred to as a
\textit{maximum independent set} of $G$. The \textit{independence number }of
$G$, denoted by $\alpha(G)$, is the cardinality of a maximum independent
set\textit{\ }of $G$.

By $\mathrm{Ind}(G)$ we mean the set of all independent sets of $G$. Let
$\Omega(G)$ denote the set of all maximum independent sets of $G$
\cite{levm3}, and
\[
\mathrm{core}(G)=\cap\{S:S\in\Omega(G)\}.
\]

A matching (i.e., a set of non-incident edges of $G$) of maximum cardinality
$\mu(G)$ is a \textit{maximum matching}, and a \textit{perfect matching} is
one covering all vertices of $G$. A vertex $v\in V(G)$ is $\mu$%
\textit{-critical} provided $\mu(G-v)<\mu(G)$.

It is well-known that
\[
\lfloor n/2\rfloor+1\leq\alpha(G)+\mu(G)\leq n
\]
hold for any graph $G$ with $n$ vertices. If $\alpha(G)+\mu(G)=n$, then $G$ is
called a \textit{K\"{o}nig-Egerv\'{a}ry graph }(a \emph{K-E }graph, for
short). We attribute this definition to Deming \cite{dem}, and Sterboul
\cite{ster}. These graphs were studied in
\cite{bourpull,korach,lov,lovpl,pulleybl}, and generalized in
\cite{bourhams1,pasdema}. Several properties of \emph{K-E} graphs are
presented in \cite{levm2,levm4,LevMan3,LevMan2006,LevMan2007}.

\begin{theorem}
\cite{levm4}\label{th2} If $G=(V,E)$ is a \textit{K\"{o}nig-Egerv\'{a}ry}%
\emph{\ }graph, then:

\emph{(i)} each maximum matching $M$ of $G$ matches $N($\textrm{core}$(G))$
into \textrm{core}$(G)$;

\emph{(ii)} $H=G-N[$\textrm{core}$(G)]$ is a \emph{K-E }graph with a perfect
matching and each maximum matching of $H$ can be enlarged to a maximum
matching of $G$.
\end{theorem}

\begin{figure}[h]
\setlength{\unitlength}{1cm}\begin{picture}(5,1.8)\thicklines
\multiput(4,0.5)(1,0){5}{\circle*{0.29}}
\multiput(5,1.5)(2,0){2}{\circle*{0.29}}
\put(4,0.5){\line(1,0){4}}
\put(5,0.5){\line(0,1){1}}
\put(7,1.5){\line(1,-1){1}}
\put(7,0.5){\line(0,1){1}}
\put(4,0.1){\makebox(0,0){$a$}}
\put(4.7,1.5){\makebox(0,0){$b$}}
\put(6,0.1){\makebox(0,0){$c$}}
\put(5,0.1){\makebox(0,0){$u$}}
\put(7,0.1){\makebox(0,0){$v$}}
\put(6.7,1.5){\makebox(0,0){$x$}}
\put(8,0.1){\makebox(0,0){$y$}}
\put(3.2,1){\makebox(0,0){$G$}}
\end{picture}
\caption{$G$ is a \emph{K-E} graph with $\alpha(G)=\left\vert \left\{
a,b,c,x\right\}  \right\vert =4$ and $\mu(G)=\left\vert \left\{
au,cv,xy\right\}  \right\vert =3$.}%
\label{fig112}%
\end{figure}
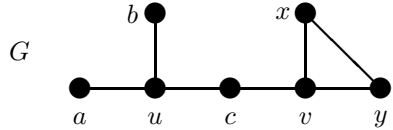

According to a well-known result of K\"{o}nig \cite{koen} and Egerv\'{a}ry
\cite{eger}, every bipartite graph is a \emph{K-E }graph. This class includes
also some non-bipartite graphs (see, for instance, the graph from Figure
\ref{fig112}).

It is easy to see that if $G$ is a \emph{K-E }graph, then $\alpha(G)\geq
\mu(G)$, and that a graph $G$ having a perfect matching is a \emph{K-E }graph
if and only if $\alpha(G)=\mu(G)$.

If $S$ is an independent set of a graph $G$ and $H=G[V-S]$, then we write
$G=S\ast H$. Clearly, any graph admits such representations. However, some
particular cases are of special interest. For instance, if $E(H)=\emptyset$,
then $G=S\ast H$ is bipartite; if $H$ is complete, then $G=S\ast H$ is a
\textit{split graph\ }\cite{FolHammer}.

\begin{proposition}
\cite{levm4}\label{prop2} If $G$ is a graph, then the following assertions are equivalent:

\emph{(i)} $G$ is a \textit{K\"{o}nig-Egerv\'{a}ry} graph;

\emph{(ii)} $G=S\ast H$, where $S\in\Omega(G)$ and $\left\vert S\right\vert
\geq\mu(G)=\left\vert V(H)\right\vert $;

\emph{(iii)} $G=S\ast H$, where $S$ is an independent set with $\left\vert
S\right\vert \geq\left\vert V(H)\right\vert $ and $(S,V(H))$ contains a
matching $M$ of size $\left\vert V(H)\right\vert $.
\end{proposition}

Let $M$ be a maximum matching of a graph $G$. To adopt Edmonds's terminology,
\cite{Edmonds}, we recall the following terms for $G$ relative to $M$. The
edges in $M$ are \textit{heavy}, while those not in $M$ are \textit{light}. An
\textit{alternating path} from a vertex $x$ to a vertex $y $ is a $x,y$-path
whose edges are alternating light and heavy. A vertex $x$ is \textit{exposed}
relative to $M$ if $x$ is not the endpoint of a heavy edge. An odd cycle $C$
with $V(C)=\{x_{0},x_{1},...,x_{2k}\}$ and
\[
E(C)=\{x_{i}x_{i+1}:0\leq i\leq2k-1\}\cup\{x_{2k},x_{0}\},
\]
such that $x_{1}x_{2},x_{3}x_{4},...,x_{2k-1}x_{2k}\in M$ is a
\textit{blossom} relative to $M$. The vertex $x_{0}$ is the \textit{base} of
the blossom. The \textit{stem} is an even length alternating path joining the
base of a blossom and an exposed vertex for $M$. The base is the only common
vertex to the blossom and the stem. A \textit{flower} is a blossom and its
stem. A \textit{posy} consists of two (not necessarily disjoint) blossoms
joined by an odd length alternating path whose first and last edges belong to
$M$. The endpoints of the path are exactly the bases of the two blossoms.

\begin{theorem}
\cite{ster}\label{th3} For a graph $G$, the following properties are equivalent:

\emph{(i)} $G$ is a \textit{K\"{o}nig-Egerv\'{a}ry graph};

\emph{(ii)} there exist no flower and no posy relative to some maximum
matching $M$;

\emph{(iii)} there exist no flower and no posy relative to every maximum
matching $M$.
\end{theorem}

\begin{figure}[h]
\setlength{\unitlength}{1cm}\begin{picture}(5,4.3)\thicklines
\multiput(2,1)(1,0){5}{\circle*{0.29}}
\multiput(2,2)(1,0){5}{\circle*{0.29}}
\put(2,1){\line(1,0){4}}
\put(2,2){\line(1,0){4}}
\multiput(2,1)(0,0.1){10}{\circle*{0.02}}
\multiput(3,1)(0,0.1){10}{\circle*{0.02}}
\multiput(4,1)(0,0.1){10}{\circle*{0.02}}
\multiput(5,1)(0,0.1){10}{\circle*{0.02}}
\multiput(6,1)(0,0.1){10}{\circle*{0.02}}
\qbezier(2,2)(4,2.8)(6,2)
\qbezier(2,1)(4,0.2)(6,1)
\put(1.8,0.6){\makebox(0,0){$1$}}
\put(2.7,1.3){\makebox(0,0){$2$}}
\put(6.4,0.6){\makebox(0,0){$2k+1$}}
\multiput(8,2)(1,0){4}{\circle*{0.29}}
\multiput(8,1)(1,0){4}{\circle*{0.29}}
\multiput(8,1)(0,0.1){10}{\circle*{0.02}}
\multiput(9,1)(0,0.1){10}{\circle*{0.02}}
\multiput(10,1)(0,0.1){10}{\circle*{0.02}}
\multiput(11,1)(0,0.1){10}{\circle*{0.02}}
\put(8,1){\line(1,0){3}}
\put(8,2){\line(1,0){3}}
\qbezier(8,2)(12.4,3.3)(11,1)
\qbezier(8,1)(12.4,-0.3)(11,2)
\put(7.8,0.6){\makebox(0,0){$1$}}
\put(9,0.6){\makebox(0,0){$2$}}
\put(11.5,0.55){\makebox(0,0){$2k$}}
\multiput(1,3)(1,0){2}{\circle*{0.29}}
\multiput(1,3)(0.1,0){10}{\circle*{0.02}}
\put(1,4){\circle*{0.29}}
\put(1,4){\line(1,-1){1}}
\put(1,3){\line(0,1){1}}
\put(0.7,4){\makebox(0,0){$v$}}
\multiput(4,3)(1,0){4}{\circle*{0.29}}
\put(4,4){\circle*{0.29}}
\put(4,3){\line(1,0){1}}
\put(6,3){\line(1,0){1}}
\put(4,4){\line(1,-1){1}}
\multiput(4,3)(0,0.1){10}{\circle*{0.02}}
\multiput(5,3)(0.1,0){10}{\circle*{0.02}}
\put(7.3,3){\makebox(0,0){$v$}}
\multiput(9,3)(1,0){4}{\circle*{0.29}}
\multiput(9,4)(3,0){2}{\circle*{0.29}}
\multiput(9,3)(0,0.1){10}{\circle*{0.02}}
\multiput(10,3)(0.1,0){10}{\circle*{0.02}}
\multiput(12,3)(0,0.1){10}{\circle*{0.02}}
\put(9,3){\line(1,0){1}}
\put(11,3){\line(1,0){1}}
\put(9,4){\line(1,-1){1}}
\put(11,3){\line(1,1){1}}
\end{picture}
\caption{Forbidden configurations. The vertex $v$ is not adjacent to the
matching edges (namely, dashed edges).}%
\label{fig444}%
\end{figure}
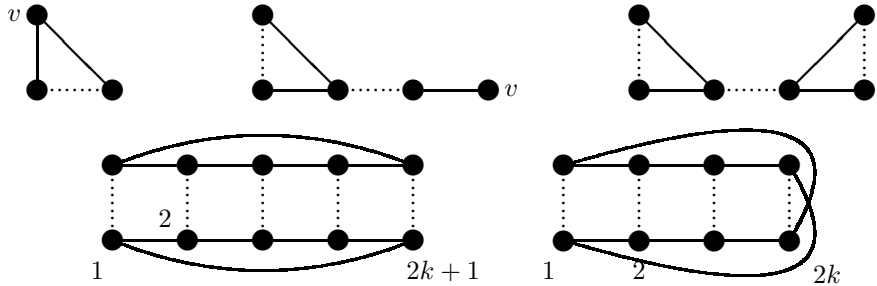

In \cite{Gavril}, Gavril defined the so-called red/blue-split graphs, as a
common generalization of \emph{K-E} and split graphs. Namely, $G$ is a
\textit{red/blue-split graph} if its edges can be colored in red and blue such
that $V(G)$ can be partitioned into a red and a blue independent set (where
\textit{red} or \textit{blue independent set} is an independent set in the
graph made of red or blue edges). In \cite{KoNgPeis}, Korach \textit{et al.}
described red/blue-split graphs in terms of excluded configurations, which led
them to the following characterization of \emph{K-E} graphs.

\begin{theorem}
\cite{KoNgPeis} Let $M$ be a maximum matching in a graph $G$. Then $G$ is a
\textit{K\"{o}nig-Egerv\'{a}ry} graph if and only if $G$ does not contain one
of the forbidden configurations, depicted in Figure \ref{fig444}, with respect
to $M$.
\end{theorem}

In \cite{lov}, Lovasz gives a characterization of \emph{K-E} graphs having a
perfect matching, in terms of certain forbidden subgraphs with respect to a
specific perfect matching of the graph.

The problem of recognizing \emph{K-E} graphs is polynomial as proved by Deming
\cite{dem}, of complexity $O(\left\vert V(G)\right\vert \left\vert
E(G)\right\vert )$. Gavril \cite{Gavril} has described a recognition algorithm
for \emph{K-E} graphs of complexity $O(\left\vert V(G)\right\vert +\left\vert
E(G)\right\vert )$. The problem of finding a maximum independent set in a
\emph{K-E} graph is polynomial as proved by Deming \cite{dem}.

The number
\[
d(G)=\max\{\left\vert S\right\vert -\left\vert N(S)\right\vert :S\in
\mathrm{Ind}(G)\}
\]
is called the \textit{critical difference} of $G$. An independent set $A$ is
\textit{critical} if $\left\vert A\right\vert -\left\vert N(A)\right\vert
=d(G)$, and the \textit{critical independence number} $\alpha_{c}(G)$ is the
cardinality of a maximum critical independent set \cite{Zhang}. Clearly,
$\alpha_{c}(G)\leq\alpha(G)$ holds for any graph $G$. It is known that the
problem of finding a critical independent set is polynomially solvable
\cite{Ageev,Zhang}.

In \cite{Larson2009} it was shown that $G$ is a \emph{K-E} graph if and only
if $\alpha_{c}(G)=\alpha(G)$, thus giving a positive answer to the Graffiti.pc
329 conjecture \cite{DeLaVina}.

The \textit{deficiency} of $G$, denoted by $def(G)$, is defined as the number
of exposed vertices relative to a maximum matching \cite{lovpl}. In other
words, $def(G)=\left\vert V\left(  G\right)  \right\vert -2\mu(G)$.

In \cite{LevMan2009} it was proven that the critical difference for a
\emph{K-E} graph $G$ is given by
\[
d(G)=\left\vert \mathrm{core}(G)\right\vert -\left\vert N(\mathrm{core}%
(G))\right\vert =\alpha(G)-\mu(G)=def(G)\text{,}%
\]
and using this finding it was demonstrated that $G$ is a \emph{K-E} graph if
and only if each of its maximum independent sets is critical.\mathstrut
\mathstrut

In this paper we give a new characterization of \emph{K-E} graphs based on
some common property of its maximum matchings, and further we use it in order
to investigate \emph{K-E} graphs in more detail.

\section{Results}

\begin{figure}[h]
\setlength{\unitlength}{1cm}\begin{picture}(5,1.2)\thicklines
\multiput(1.5,0)(1,0){3}{\circle*{0.29}}
\put(1.5,1){\circle*{0.29}}
\put(2.5,1){\circle*{0.29}}
\put(1.5,0){\line(1,0){2}}
\put(2.5,0){\line(0,1){1}}
\put(2.5,0){\line(-1,1){1}}
\put(3.5,0){\line(-1,1){1}}
\put(1.2,0){\makebox(0,0){$a$}}
\put(1.2,1){\makebox(0,0){$b$}}
\put(2.8,0.2){\makebox(0,0){$c$}}
\put(2.8,1){\makebox(0,0){$x$}}
\put(3.8,0){\makebox(0,0){$y$}}
\put(0.5,0.5){\makebox(0,0){$G_{1}$}}
\multiput(5.5,0)(1,0){4}{\circle*{0.29}}
\put(7.5,1){\circle*{0.29}}
\put(6.5,1){\circle*{0.29}}
\put(5.5,0){\line(1,0){3}}
\put(6.5,1){\line(1,0){1}}
\put(7.5,0){\line(0,1){1}}
\put(6.5,0){\line(1,1){1}}
\put(5,0.5){\makebox(0,0){$G_{2}$}}
\multiput(10.5,0)(1,0){3}{\circle*{0.29}}
\multiput(10.5,1)(1,0){2}{\circle*{0.29}}
\put(10.5,0){\line(1,0){2}}
\put(10.5,0){\line(0,1){1}}
\put(11.5,0){\line(0,1){1}}
\put(11.5,1){\line(1,-1){1}}
\put(10.2,1){\makebox(0,0){$u$}}
\put(11.8,1){\makebox(0,0){$v$}}
\put(10.2,0){\makebox(0,0){$x$}}
\put(11.23,0.25){\makebox(0,0){$z$}}
\put(12.8,0){\makebox(0,0){$y$}}
\put(9.7,0.5){\makebox(0,0){$G_{3}$}}
\end{picture}
\caption{$G_{1}$ and $G_{2}$ are K\"{o}nig--Egerv\'{a}ry graphs, but only in
$G_{2}$ has a perfect matching. $G_{3}$ is not a K\"{o}nig--Egerv\'{a}ry
graph.}%
\label{fig22}%
\end{figure}
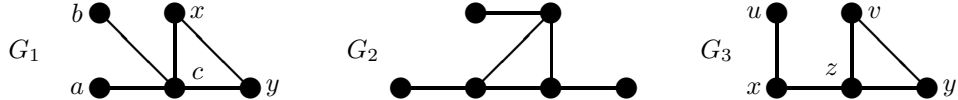

Notice that all the maximum matchings of the graphs $G_{1}$ and $G_{2}$ from
Figure \ref{fig22} are included in $(S,V(G_{i})-S),i=1,2$, for each
$S\in\Omega(G_{i}),i=1,2$. On the other hand, $M_{1}=\{xu,yz\}$ and
$M_{2}=\{xu,vz\}$ are maximum matchings of the graph $G_{3}$ from Figure
\ref{fig22}, and $S=\{u,v\}\in\Omega(H_{2})$, but $M_{1}\nsubseteq
(S,V(G_{3})-S)$, while $M_{2}\subseteq(S,V(G_{3})-S)$. 

\begin{theorem}
\label{th1}For a graph $G=(V,E)$, the following properties are equivalent:

\emph{(i)} $G$ is a \textit{K\"{o}nig-Egerv\'{a}ry graph};

\emph{(ii)} each maximum matching of $G$ is contained in $(S,V-S)$ for some
$S\in\Omega(G)$;

\emph{(iii) }each maximum matching of $G$ is contained in $(S,V-S)$ for every
$S\in\Omega(G)$.
\end{theorem}

\begin{proof}
\emph{(i) }$\Longrightarrow$ \emph{(iii) }Let $G$ be a \emph{K-E} graph.
Suppose that there exist some $S\in\Omega(G)$ and a maximum matching $M$ such
that $M\nsubseteq(S,V-S)$. According to Proposition \ref{prop2}, $G$ can be
written as $G=S\ast H$ and $\mu(G)=\left\vert V(H)\right\vert =\left\vert
V-S\right\vert $. Since $S$ is independent and $M\nsubseteq(S,V-S)$, there
must be an edge in $M\cap E(H)$. Hence, we infer that $\mu(G)<\left\vert
V(H)\right\vert $, in contradiction with $\mu(G)=\left\vert V(H)\right\vert $.
Therefore, $M$ must be contained in $(S,V-S)$.

\emph{(iii) }$\Longrightarrow$ \emph{(ii) }It is clear.

\emph{(ii) }$\Longrightarrow$ \emph{(i) }Let $S\in\Omega(G)$ enjoy the
property that each maximum matching of $G$ is contained in $(S,V-S)$.

Assume, on the contrary, that $G$ is not a \emph{K-E }graph, i.e.,
$\alpha(G)+\mu(G)<\left\vert V(G)\right\vert $. Let $M=\{a_{k}b_{k}:1\leq
k\leq\mu(G)\}$ be a maximum matching in $G$. Since $M\subseteq(S,V-S)$, we
infer that $\mu(G)\leq\left\vert S\right\vert =\alpha(G)$, and one may suppose
that
\begin{align*}
A  &  =\{a_{k}:1\leq k\leq\mu(G)\}\subseteq S\text{, while}\\
B  &  =\{b_{k}:1\leq k\leq\mu(G)\}\subseteq V-S.
\end{align*}
In addition, it follows that $\mu(G)<\left\vert V-S\right\vert $, because
\[
\left\vert S\right\vert +\left\vert M\right\vert =\alpha(G)+\mu(G)<\left\vert
V\right\vert =\left\vert S\right\vert +\left\vert V-S\right\vert
=\alpha(G)+\left\vert V-S\right\vert .
\]

Let $x\in V-S-B$ and $S_{x}$ be the set of vertices $v\in B$ such that there
exists a path $x=v_{1},v_{2},...,v_{2k+1}=v$, where $v_{2i}v_{2i+1}\in
M,v_{2i}\in A$ and $v_{2i+1}\in B$. We show that the set $S_{1}=\{x\}\cup
S_{x}\cup(S-M(S_{x}))$ is independent, where $M(S_{x})=\{a_{j}\in A:b_{j}\in
S_{x}\}$.

\emph{Claim 1}. $\{x\}\cup(S-M(S_{x}))$ is an independent set in $G$.

Clearly, $S-M(S_{x})$ is independent, as a subset of $S$. In addition, if
$xy\in E$, for some $y\in S-M(S_{x})$, then, according to the definition of
$S_{x}$, no edge issuing from $y$ belongs to $M$. Hence, $M\cup\{xy\}$ is a
matching in $G$, larger than $M$, in contradiction to the maximality of $M$.
Therefore, $\{x\}\cup(S-M(S_{x}))$ is independent.

\emph{Claim 2}. $S_{x}$ is independent.

Otherwise, assume that $b_{j}b_{k}\in E$ for some $b_{j},b_{k}\in S_{x}$. By
definition of $S_{x}$, there are two paths:
\[
P_{1}:x=v_{1},v_{2},...,v_{2p+1}=b_{j},
\]
where $v_{2i}v_{2i+1}\in M$, $v_{2i}\in A$ and $v_{2i+1}\in B$, and
\[
P_{2}:x=u_{1},u_{2},...,u_{2q+1}=b_{k},
\]
where $u_{2i}u_{2i+1}\in M$, $u_{2i}\in A$ and $u_{2i+1}\in B$.

\textit{Case 1}. $b_{k}=v_{2s+1}$ is on the path $P_{1}$ (similarly, when
$b_{j}=u_{2s+1}$ on the path $P_{2}$).

Then, it follows that
\[
M_{1}=\{v_{1}v_{2},v_{3}v_{4},...,v_{2s-1}v_{2s}\}\cup\{v_{2s+3}%
v_{2s+4},v_{2s+5}v_{2s+6},...,v_{2p-1}v_{2p}\}\cup\{b_{j}b_{k}\}
\]
is a matching with $p$ edges, and
\[
M_{2}=M_{1}\cup(M-\{v_{2i}v_{2i+1}:1\leq i\leq p\})
\]
is a maximum matching in $G$. This contradicts the assumption that
$M_{2}\subseteq(S,V-S)$, because $b_{j},b_{k}\in S_{x}\subseteq V-S$.

\textit{Case 2}. The paths $P_{1}$ and $P_{2}$ have in common only the vertex
$x$.

The edge $b_{j}b_{k}$ closes a cycle with he paths $P_{1}$ and $P_{2}$. Now,
the sets%
\[
M_{3}=\{v_{1}v_{2},v_{3}v_{4},...,v_{2p-1}v_{2p}\}\cup\{u_{3}u_{4},u_{5}%
u_{6},...,u_{2q-1}u_{2q}\}\cup\{b_{j}b_{k}\},
\]
and
\[
M_{4}=\{v_{2i}v_{2i+1}:1\leq i\leq p\}\cup\{u_{2i}u_{2i+1}:1\leq i\leq q\}
\]
are disjoint matchings in $G$, both with $p+q$ edges, while $M_{5}=M\cup
M_{3}-M_{4}$ is a maximum matching that satisfies $M_{5}\nsubseteq(S,V-S)$, in
contradiction to the hypothesis.

Therefore, $S_{x}$ must be an independent set in $G$.

\emph{Claim 3}. No edge joins $x$ to some vertex of $S_{x}$.

Suppose, on the contrary, that there is $b_{j}\in S_{x}$, such that $xb_{j}\in
E$. By the definition of $S_{x}$, there is a path $x=v_{1},v_{2}%
,...,v_{2p+1}=b_{j}$, where $v_{2i}v_{2i+1}\in M,v_{2i}\in A$ and $v_{2i+1}\in
B$. Then $M_{1}=\{v_{1}v_{2},v_{3}v_{4},...,v_{2p-1}v_{2p}\} $ is a matching
in $G$ with $p$ edges, and
\[
M_{2}=M\cup M_{1}\cup\{xb_{j}\}-\{v_{2i}v_{2i+1}:1\leq i\leq p\}
\]
is a maximum matching of $G$. Since $x,b_{j}\in V-S$, it follows that
$M_{2}\nsubseteq(S,V-S)$, again in contradiction to the hypothesis.

\emph{Claim 4}. No edge joins a vertex from $S-M(S_{x})$ to a vertex of
$S_{x}$.

Otherwise, assume that there is $y\in S-M(S_{x}),b_{j}\in S_{x}$, such that
$xb_{j}\in E$. As above, there is a path $x=v_{1},v_{2},...,v_{2p+1}=b_{j}$,
where $v_{2i}v_{2i+1}\in M,v_{2i}\in A$ and $v_{2i+1}\in B$. Then, the set
$M_{1}=\{v_{1}v_{2},v_{3}v_{4},...,v_{2p-1}v_{2p}\}$ is a matching in $G$ with
$p$ edges, and
\[
M_{2}=M\cup M_{1}\cup\{yb_{j}\}-\{v_{2i}v_{2i+1}:1\leq i\leq p\}
\]
is a matching of $G$ larger than $M$, thus contradicting the maximality of $M
$.

Finally, we may conclude that
\[
S_{1}=\{x\}\cup S_{x}\cup(S-M(S_{x}))
\]
is an independent set in $G$, but this leads to the following inequality
\[
\left\vert S_{1}\right\vert =\left\vert S\right\vert +1>\alpha(G),
\]
which clearly contradicts the fact that $\alpha(G)$ is the size of a maximum
independent set in $G$.
\end{proof}

\begin{proposition}
\label{prop3}If $G=(V,E)$ is a \textit{K\"{o}nig-Egerv\'{a}ry}\emph{\ }graph, then

\emph{(i)} for every maximum matching each exposed vertex belongs to
\textrm{core}$(G)$.

\emph{(ii)} at least one of the endpoints of every edge of $G$ is a $\mu
$-critical vertex.
\end{proposition}

\begin{proof}
\emph{(i)} By Theorem \ref{th1}, every maximum matching $M$ is included in
$(S,V-S)$, for each maximum independent set $S$. Since $\left\vert
M\right\vert =\left\vert V-S\right\vert $, we deduce that no exposed vertex
belongs to $V-S$, and consequently, no exposed vertex is in
\[
\cup\{V-S:S\in\Omega(G)\}=V-\cap\{S:S\in\Omega(G)\}=V-\text{\textrm{core}%
}(G).
\]
In other words, every exposed vertex belongs to \textrm{core}$(G)$.

\emph{(ii)} Suppose that $uv\in E$ and $v$ is not $\mu$-critical, i.e.,
$\mu(G-v)=\mu(G)$.

If $\alpha(G-v)=\alpha(G)$, then we get the following contradiction:
\[
\left\vert V\right\vert -1\geq\alpha(G-v)+\mu(G-v)=\alpha(G)+\mu(G)=\left\vert
V\right\vert .
\]
Therefore, we infer that $\alpha(G-v)=\alpha(G)-1$, i.e., $v\in$
\textrm{core}$(G)$. Hence, $u\in N($\textrm{core}$(G))$, and, consequently,
$u$ is $\mu$-critical, because $N($\textrm{core}$(G))$ is matched into
$\mathrm{core}(G)$ by every maximum matching in a \emph{K-E} graph (by Theorem
\ref{th2}\emph{(i)}).
\end{proof}

\begin{figure}[h]
\setlength{\unitlength}{1cm}\begin{picture}(5,1)\thicklines
\multiput(2.5,0)(1,0){4}{\circle*{0.29}}
\put(4.5,1){\circle*{0.29}}
\put(3.5,1){\circle*{0.29}}
\put(3.5,0){\line(0,1){1}}
\put(2.5,0){\line(1,0){3}}
\put(4.5,0){\line(0,1){1}}
\put(5.5,0){\line(-1,1){1}}
\put(2.5,0){\line(1,1){1}}
\put(1.8,0.5){\makebox(0,0){$W$}}
\multiput(8,1)(1,0){4}{\circle*{0.29}}
\multiput(9,0)(2,0){2}{\circle*{0.29}}
\put(8,0){\circle*{0.29}}
\put(8,0){\line(1,0){3}}
\put(8,1){\line(1,-1){1}}
\put(9,0){\line(0,1){1}}
\put(9,0){\line(1,1){1}}
\put(9,0){\line(2,1){2}}
\put(9,1){\line(2,-1){2}}
\put(9,1){\line(1,0){2}}
\put(10,1){\line(1,-1){1}}
\put(11,0){\line(0,1){1}}
\qbezier(9,1)(10,1.7)(11,1)
\put(7.2,0.5){\makebox(0,0){$H$}}
\end{picture}
\caption{The non-K\"{o}nig-Egerv\'{a}ry graphs $W$ and $H$ have all exposed
vertices in $core(W)$ and $core(H)$, respectively.}%
\label{fig33}%
\end{figure}
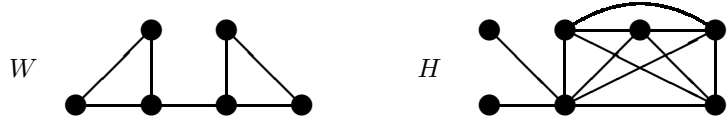

\begin{remark}
The converse of Proposition \ref{prop3}\emph{(i)} is false (see the graphs in
Figure \ref{fig33}).
\end{remark}

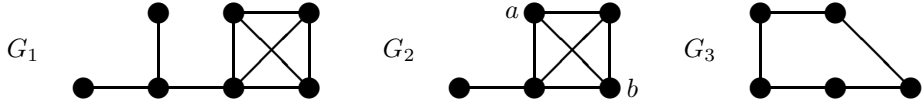
\begin{figure}[h]
\setlength{\unitlength}{1cm}\begin{picture}(5,1)\thicklines
\multiput(2,0)(1,0){4}{\circle*{0.29}}
\multiput(3,1)(1,0){3}{\circle*{0.29}}
\put(2,0){\line(1,0){3}}
\put(3,0){\line(0,1){1}}
\put(4,0){\line(0,1){1}}
\put(4,0){\line(1,1){1}}
\put(4,1){\line(1,-1){1}}
\put(4,1){\line(1,0){1}}
\put(5,0){\line(0,1){1}}
\put(1.2,0.5){\makebox(0,0){$G_{1}$}}
\multiput(7,0)(1,0){3}{\circle*{0.29}}
\multiput(8,1)(1,0){2}{\circle*{0.29}}
\put(7,0){\line(1,0){2}}
\put(8,0){\line(0,1){1}}
\put(8,0){\line(1,1){1}}
\put(8,1){\line(1,-1){1}}
\put(8,1){\line(1,0){1}}
\put(9,0){\line(0,1){1}}
\put(7.7,1){\makebox(0,0){$a$}}
\put(9.3,0){\makebox(0,0){$b$}}
\put(6.2,0.5){\makebox(0,0){$G_{2}$}}
\multiput(11,0)(1,0){3}{\circle*{0.29}}
\multiput(11,1)(1,0){2}{\circle*{0.29}}
\put(11,0){\line(1,0){2}}
\put(11,0){\line(0,1){1}}
\put(11,1){\line(1,0){1}}
\put(12,1){\line(1,-1){1}}
\put(10.2,0.5){\makebox(0,0){$G_{3}$}}
\end{picture}
\caption{All $G_{i}$, $i=1,2,3$, are not K\"{o}nig--Egerv\'{a}ry graphs. }%
\label{fig222}%
\end{figure}

\begin{remark}
Proposition \ref{prop3}\emph{(ii)} is not specific for \emph{K-E} graphs; see,
for instance, the graph $G_{1}$ from Figure \ref{fig222}. On the other hand,
there exist graphs where the endpoints of \emph{(a)} some edges are not $\mu
$-critical (e.g., the edge $ab$ of the graph $G_{2}$ from Figure
\ref{fig222}), \emph{(b)} each edge are not $\mu$-critical (e.g., the graph
$G_{3}$ from Figure \ref{fig222}).
\end{remark}

\begin{proposition}
Let $G$ be a \textit{K\"{o}nig-Egerv\'{a}ry} graph $G$ and $v\in V(G)$ be such
that $G-v$ is still a \textit{K\"{o}nig-Egerv\'{a}ry} graph. Then
$v\in\mathrm{core}(G)$ if and only if there exists a maximum matching that
does not saturate $v$.
\end{proposition}

\begin{proof}
Since $v\in\mathrm{core}(G)$, it follows that $\alpha(G-v)=\alpha(G)-1$.
Consequently, we have
\[
\alpha(G)+\mu(G)-1=\left\vert V(G)\right\vert -1=\left\vert V(G-v)\right\vert
=\alpha(G-v)+\mu(G-v)
\]
which implies that $\mu(G)=\mu(G-v)$. In other words, there is a maximum
matching in $G$ not saturating $v$.

Conversely, suppose that there exists a maximum matching in $G$ that does not
saturate $v$. Since, by Theorem \ref{th2}\emph{(i)}, $N($\textrm{core}$(G))$
is matched into \textrm{core}$(G)$ by every maximum matching, it follows that
$v\notin N($\textrm{core}$(G))$.

Assume that $v\notin$ \textrm{core}$(G)$. By Theorem \ref{th2}\emph{(ii)},
$H=G-N\left[  \mathrm{core}(G)\right]  $ is a \emph{K-E} graph, $H$ has a
perfect matching and every maximum matching $M$ of $G$ is of the form
$M=M_{1}\cup M_{2}$, where $M_{1}$ matches $N($\textrm{core}$(G))$ into
\textrm{core}$(G)$, while $M_{2}$ is a perfect matching of $H$. Consequently,
$v$ is saturated by every maximum matching of $G$, in contradiction with the
hypothesis on $v$.
\end{proof}

\begin{remark}
The above proposition is not true if $G-v$ is not a \emph{K-E} graph; e.g.,
each maximum matching of the graph $G$ from Figure \ref{fig112} saturates
$c\in$ \textrm{core}$(G)=\{a,b,c\}$.
\end{remark}

\begin{corollary}
For every bipartite graph $G$, the vertex $v\in\mathrm{core}(G)$ if and only
if there exists a maximum matching that does not saturate $v$.
\end{corollary}

\section{Conclusions}

In this paper we give a new characterization of K\"{o}nig-Egerv\'{a}ry graphs
similar in form to Sterboul's Theorem \ref{th3}. It seems to be interesting to
characterize K\"{o}nig-Egerv\'{a}ry graphs with unique maximum independent sets.

\end{document}